\documentclass[cameraready]{Interspeech}
\DeclareUnicodeCharacter{2060}{}

\title{An auscultation location specific study on the relationship between expiratory-to-inspiratory acoustic patterns and spirometric airflow limitation across age and gender in asthmatic patients}

\author[affiliation={1},orcid=0009-0000-8757-8515]{Dheeraj}{Harish Kumar}
\author[affiliation={1},orcid=0009-0005-6028-4521]{Sanjana}{MC}
\author[affiliation={1}]{Perumal}{Keerthi Priya}
\author[affiliation={1}]{K V}{Nikhath Khanam}
\author[affiliation={2},orcid=0000-0002-5144-4731]{Uma Maheshwari}{Krishnaswamy}
\author[affiliation={1},orcid=0000-0002-1137-0838]{Prasanta}{Kumar Ghosh}

\address{
    $^1$ Indian Institute of Science, Bangalore, India 
    $^2$ St. Johns National Academy of Health Sciences, Bangalore-560034, India 
}

\email{dheerajh@iisc.ac.in, sanjanac@iisc.ac.in, keerthipriya.spirelab.iisc@gmail.com nikhathk@iisc.ac.in, umamohan99@gmail.com, prasantg@iisc.ac.in}

\keywords{}

\usepackage{comment}
\usepackage{booktabs}
\usepackage{longtable}

\begin{document}

\maketitle

\begin{abstract}
\vspace{-1ex}
Asthma causes expiratory airflow limitation and is clinically assessed using spirometry, which provides the FEV$_1$/FVC ratio representing the proportion of air exhaled in the first second relative to total forced vital capacity. Prior studies suggest that respiratory sounds recorded at posterior sites (Left Lower, Left Upper, Right Upper, Right Lower) reflect regional airflow patterns. In this study, we investigate the relationship between the expiratory-to-inspiratory (E/I) spectral power ratio and FEV$_1$/FVC in 141 participants aged 20-60 years using Spearman correlation across frequency subbands. The 100-200 Hz and 200-400 Hz bands showed significant correlations. Overall, lower posterior sites showed stronger associations; younger adults showed stronger correlations at the Left Lower site, whereas older adults showed stronger correlations at the Left Upper site. Gender-stratified analysis showed stronger Left Lower correlations in males and stronger Left Upper correlations in females.
\end{abstract}
Asthma, Lung Sound Analysis, Expiratory-to-Inspiratory Ratio, Spirometry, Respiratory Acoustics, Acoustic Biomarkers.

\vspace{-1.5ex}
\section{Introduction}
\vspace{-1ex}
Asthma is a chronic inflammatory disease of the airways resulting in symptoms such as bronchial obstruction, chest discomfort, and wheezing \cite{martinez2006genes,Zhang2025}. Affecting approximately 262 million people globally, it remains a significant public health burden with nearly 455,000 annual deaths worldwide \cite{Zhang2025}. The gold standard for monitoring asthma severity is spirometry. During spirometry, participants are instructed to wear a nose clip, take a deep breath to their maximum capacity, and then exhale into the spirometer as forcefully and as long as possible, preferably for at least six seconds. This maneuver primarily depends on the participant’s effort and cooperation, causing spirometry readings to vary depending on how accurately the inhalation and exhalation are performed. It becomes difficult to obtain reliable spirometry readings in children and elderly individuals \cite{pellegrino2005interpretative}, as the procedure can be strenuous and challenging to perform correctly.

Spirometry provides key parameters such as Forced Expiratory Volume in one second (FEV$_1$), Forced Vital Capacity (FVC), and the FEV$_1$/FVC ratio. FEV$_1$ refers to the volume of air that can be forcibly exhaled during the first second of a maximal expiration and reflects the degree of airway obstruction. FVC represents the total volume of air that can be forcefully exhaled after full inspiration, indicating overall ventilatory capacity. The FEV$_1$/FVC ratio expresses the proportion of air expelled in the first second relative to the total forced vital capacity and is widely used to identify airflow limitation; reduced values typically indicate obstructive airway disease \cite{miller2005standardisation}. In obstructive airway diseases such as asthma, airflow obstruction predominantly affects the expiratory phase because as the lungs deflate during breathing out, the small airways tend to narrow more easily and may partially collapse, increasing resistance to airflow. In contrast, during inspiration, lung expansion helps keep the airways open. This primary limitation of expiratory flow leads to a greater reduction in FEV$_1$ and consequently decreases the FEV$_1$/FVC ratio \cite{miller2005standardisation, calverley2005flow}.

Exhaled spectral power (E) refers to the average of the squared magnitude of the short-time Fourier transform computed over the expiratory segment and represents the mean acoustic power generated during breathing out \cite{pasterkamp1997respiratory}. Similarly, inhaled spectral power (I) is the average over the inspiratory segment and reflects the mean acoustic power during breathing in \cite{pasterkamp1997respiratory}. The expiratory-to-inspiratory (E/I) ratio reflects how dominant expiratory sound power is compared to inspiratory sound power. Since spirometric indices such as FEV$_1$ and FEV$_1$/FVC measure expiratory airflow, and obstructive airway diseases primarily affect expiration, examining expiratory acoustic characteristics provides a physiologically consistent basis for comparison with spirometric airflow limitation.

Since both E/I and FEV$_1$/FVC are influenced by expiratory airflow mechanics, examining their correlation helps determine whether changes in acoustic patterns truly reflect underlying physiological airflow limitation \cite{mineshita2014correlation}. If a meaningful relationship exists, it suggests that the E/I ratio could serve as a non-invasive and less effort-dependent marker of obstruction severity in asthma \cite{shimoda2017lung}. Several studies have consistently reported meaningful associations between E/I spectral measures and spirometric indices of airflow limitation \cite{mineshita2014correlation, shimoda2017lung, shimoda2017lungf, karimizadeh2021multichannel, thap2016high}. Shimoda et al. \cite{shimoda2017lung, shimoda2017lungf} examined 84 and 22 participants with bronchial asthma in two separate studies, respectively. In the study involving 84 patients, mid-frequency E/I ratios (100-400 Hz) were significantly correlated with FEV$_1$ and FEV$_1$/FVC and showed improvement following therapeutic intervention. In the smaller study of 22 patients, mid-frequency E/I ratios were again significantly correlated with FEV$_1$/FVC, with the strongest associations observed at the left anterior and left posterior lower lung regions. Similar relationships have been observed in Chronic Obstructive Pulmonary Disease (COPD) populations. Mineshita et al. \cite{mineshita2014correlation} conducted a study in 39 COPD patients and demonstrated that the lower-to-upper lung sound intensity ratio (Lower QLD/Upper QLD) was significantly associated with FEV$_1$\% predicted and FEV$_1$/FVC, with increasing obstruction corresponding to greater upper-lung sound dominance. Beyond asthma and COPD, Karimizadeh et al. \cite{karimizadeh2021multichannel} analyzed 209 multichannel recordings from 37 cystic fibrosis patients and showed that expiration-to-inspiration power ratios from large, upper, and peripheral airway regions distinguished disease severity, with region-specific features reflecting progression. Thap et al. \cite{thap2016high} examined 26 subjects (13 healthy individuals and 13 COPD patients) and reported that time-frequency analysis of forced expiratory sounds was strongly associated with FEV$_1$/FVC (r = 0.814), whereas FEV$_1$ and FVC showed weaker relationships, suggesting that acoustic analysis may provide useful information related to airflow limitation.

Previous studies have examined acoustic E/I measures across respiratory conditions such as asthma, COPD, and cystic fibrosis, analyzing variations across different frequency bands and auscultation locations. However, the potential influence of demographic factors such as age and gender on location-specific acoustic characteristics remains underexplored. With aging, reduced elastic recoil and increased closing volume lead to earlier small airway closure, particularly in the lower lung regions \cite{babb2000mechanism, turner1968elasticity}. This may reduce airflow contribution from the lung bases and alter the spatial distribution of lung sound energy. In addition, anatomical differences between males and females including variations in airway diameter, lung size, and lung volume \cite{bellemare2003sex, dominelli2018sex, kim2011gender} may influence airflow distribution and sound transmission between males and females.

In this work, we investigate whether the relationship between the expiratory-to-inspiratory (E/I) spectral power ratio and spirometric airflow limitation exhibits age and gender dependent patterns in patients with asthma. Participants are divided into four age groups (20-30, 30-40, 40-50, and 50-60 years) and by gender (male and female). We evaluate the correlation between E/I and FEV$_1$/FVC across multiple posterior auscultation locations (Left Lower, Left Upper, Right Upper, and Right Lower) and frequency bands (0-800 Hz broadband, 100-200 Hz, 200-400 Hz, and 400-800 Hz). By examining age and gender specific variations in these acoustic-spirometric relationships, this study aims to determine whether the spatial distribution of the E/I association with FEV$_1$/FVC demonstrates systematic trends with respect to auscultation locations within demographic groups.
\begin{table}[t]
\caption{Distribution of participants across age groups and gender, along with mean FEV$_1$/FVC (\%) and total number of breath cycles.}
\centering
\small
\setlength{\tabcolsep}{4pt}
\label{tab:summary}
\begin{tabular}{lcccc}
\toprule
Age & Total (Male + Female) &  FEV$_1$/FVC (\%) & Cycles \\
\midrule
20--30 & 38 (20 + 18) & 83.30 $\pm$ 8.67  & 760\\
30--40 & 42 (18 + 24) & 83.57 $\pm$ 9.60  & 840\\
40--50 & 43 (21 + 22) & 79.37 $\pm$ 12.13 & 860\\
50--60 & 18 (7  + 11) & 86.33 $\pm$ 7.40  & 355\\
\midrule
\textbf{Total} & \textbf{141 (66 + 75) } & \textbf{82.58 $\pm$ 10.13} & \textbf{2815}\\
\bottomrule
\end{tabular}
\end{table}
\vspace{-1.5ex}
\section{Dataset}
\vspace{-1ex}
The dataset used in this study was collected from St. John’s National Academy of Health
Sciences, Bangalore and consists of 141 clinically diagnosed asthma participants, including 66 males and 75 females. The participants were between 20 and 60 years of age, with an average age of 37 ± 10.1 years. All participants were confirmed to have asthma through spirometry and the results were clinically verified by the doctor. The FEV$_1$/FVC ratio derived from spirometry was used as the primary clinical measure of asthma severity. The recorded FEV$_1$/FVC values ranged between 51\% and 99\%, indicating varying levels of asthma severity, including mild, moderate, and severe cases. The age-wise distribution of participants is summarized in Table~\ref{tab:summary}. To assess whether spirometric severity differed across age groups, pairwise (all possible pairs of age-groups) Welch's t-tests were performed on FEV$_1$/FVC values, and the corresponding results are provided in Appendix Table~\ref{tab:welch}.

Recordings were conducted under the doctor’s guidance. Prior approval for the recordings was obtained, and informed consent was obtained from each subject. Five breath cycles were recorded from each of the four posterior chest auscultation locations, namely Left Upper (LU), Left Lower (LL), Right Lower (RL), and Right Upper (RU) \cite{pasterkamp1997respiratory}, using a Littmann CORE digital stethoscope \cite{littmann_core_2021} at a sampling rate of 4 kHz. However, for one participant in 50-60 age group, the Right Upper (RU) recording was unavailable due to faulty recording.  All recordings were performed while the participants were seated in an upright position and instructed to breathe normally. The inhalation and exhalation segment boundaries were manually annotated by experts through auditory inspection and visual waveform analysis using Audacity \cite{mazzoni2000audacity}. The mean duration of inhalation segments was 1.067 ± 0.539 seconds, ranging from 0.255 seconds to 4.415 seconds. The mean duration of exhalation segments was 1.401 ± 0.795 seconds, ranging from 0.361 seconds to 7.322 seconds.

\begin{figure}[t]
    \centering
    \includegraphics[width=\linewidth]{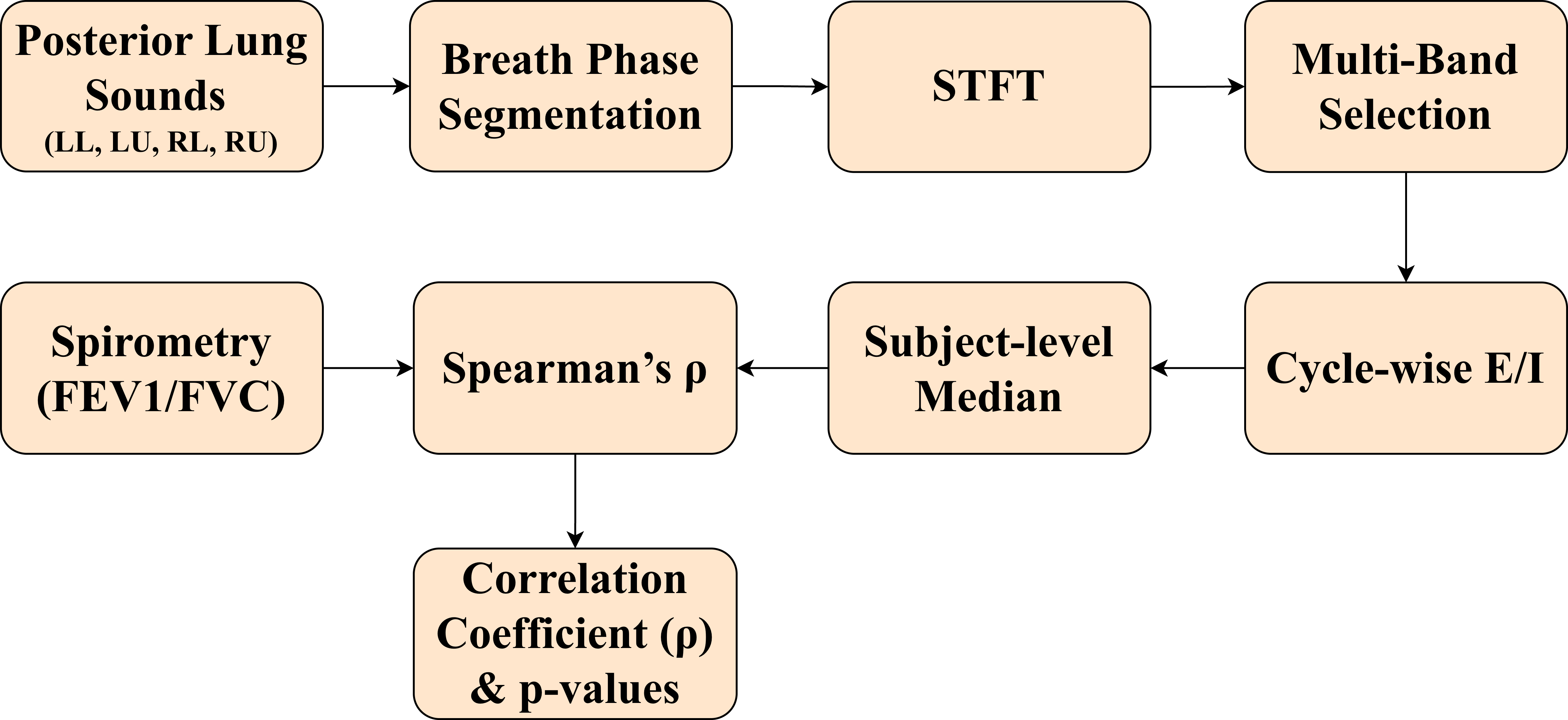}
    \caption{Block diagram of the proposed framework.}
    \label{fig:block}
\end{figure}
\begin{figure*}[t]
    \centering
    \includegraphics[width=\textwidth]{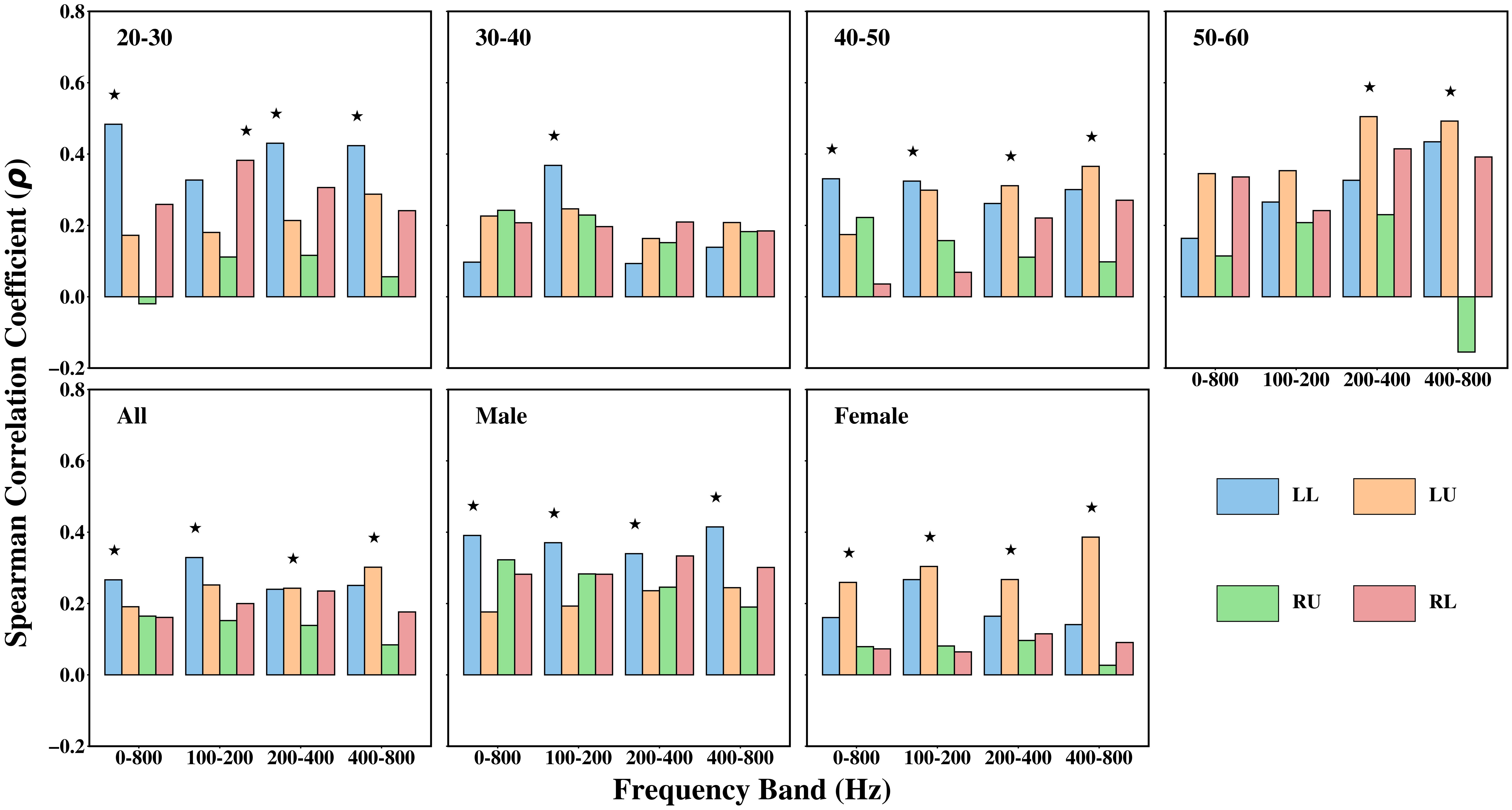}

    \caption{Age-wise and Gender-wise Spearman correlation ($\rho$) between E/I ratio and FEV$_1$/FVC across posterior auscultation locations and frequency bands ($\star$: highest correlation with $p$ $<$ 0.05).}
    \label{fig:correlation_results}
\end{figure*}
\vspace{-1.5ex}
\section{Study Design}
\vspace{-1ex}
\subsection{Setup}
\vspace{-1ex}
The overall processing and analysis pipeline is illustrated in Fig.~\ref{fig:block}. 
For each participant, respiratory sounds were recorded at four auscultation locations, 
with five breath cycles available per location. Breath phases were segmented into 
inspiration and expiration. The expiratory to inspiratory (E/I) ratio was computed 
for each segmented inspiratory and expiratory phase within a breath cycle using the 
Short Time Fourier Transform (STFT).

STFT was applied to obtain the time-frequency representation of the lung sound 
signals. Since lung sounds are non-stationary and vary dynamically during breathing, 
STFT analyzes the signal over short overlapping windows to capture time varying 
spectral characteristics and compute spectral power over small time segments before 
averaging. This enables reliable estimation of band-specific energy.

The time-frequency spectral power for respiratory phase $u$, where 
$u \in \{I,E\}$ denotes inspiration ($I$) or expiration ($E$), was computed 
as the squared magnitude of the STFT coefficients:
\vspace{-0.5ex}
\begin{equation}
P^{(u)}(n,k) =
\left|
\sum_{m=0}^{N-1}
x_u[nH + m] \, w[m] \,
e^{-j \frac{2\pi k m}{N}}
\right|^2
\end{equation}

where $x_u[\cdot]$ represents the lung sound signal corresponding to respiratory 
phase $u$, $w[m]$ is a Hanning window of length $N=2048$, $H=512$ denotes the hop 
length (frame shift) between consecutive analysis windows, $m$ is the sample index 
within the window, $n$ denotes the time frame index, and $k$ represents the frequency 
bin index. The window length of 2048 samples was chosen to provide sufficient 
frequency resolution for analyzing lung sound components below 800 Hz while 
maintaining adequate temporal resolution. The Hanning window reduces spectral 
leakage by smoothly tapering the signal at the window boundaries.

Band-limited spectral power was computed over four predefined frequency ranges: 
100-200 Hz, 200-400 Hz, 400-800 Hz, and 0-800 Hz. Frequency bins corresponding 
to each band were determined from the STFT frequency resolution, where the center 
frequency of the $k$-th bin is given by
\vspace{-1ex}
\begin{equation}
f_k = \frac{kF_s}{N} 
\end{equation}

with $F_s$ denoting the sampling frequency and $N$ the FFT length. Only bins whose 
center frequencies fell within each band were retained to compute the band-limited 
spectral power.

To obtain a representative spectral power value for each respiratory phase $u$, 
power values were averaged across time frames and the selected frequency bins:
\vspace{-1ex}
\begin{equation}
P_B^{(u)} =
\frac{1}{T K}
\sum_{n=1}^{T}
\sum_{k \in B}
P^{(u)}(n,k)
\end{equation}
where $T$ denotes the total number of time frames, $K$ represents the number of 
frequency bins within the selected band $B$, and $P^{(u)}(n,k)$ is the spectral 
power at time frame $n$ and frequency bin $k$ for respiratory phase $u$. This yields 
a single representative spectral power value for each inspiratory 
$(P_B^{(I)})$ and expiratory $(P_B^{(E)})$ segment.

Finally, the expiratory-to-inspiratory ratio was computed for each breath cycle and 
frequency band as
\vspace{-1ex}
\begin{equation}
E/I =
\frac{P_B^{(E)}}{P_B^{(I)}}
\end{equation}

To obtain a stable subject-level value at each auscultation site, the median of the five E/I values was used as the representative value. The median was preferred over the mean because individual breaths may be affected by posture changes, movement artifacts, or irregular breathing patterns.

Spearman’s rank correlation coefficient was computed to assess the monotonic relationship between E/I and FEV$_1$/FVC, as it does not assume a linear relationship or does not require the variables to be normally distributed. The analysis was performed separately for each auscultation location and frequency band, and further conducted independently within each age and gender group.

The null hypothesis ($H_0$) stated that no monotonic relationship exists between E/I and FEV$_1$/FVC. For each test, the correlation coefficient ($\rho$) and the corresponding $p$-value were reported to evaluate statistical significance. A significance level of $\alpha = 0.05$ was adopted for all statistical tests. Additional significance maps before and after Benjamini--Hochberg FDR correction are provided in Appendix Figures~\ref{fig:without_fdr} and~\ref{fig:with_fdr}, respectively.
\vspace{-1.5ex}
\subsection{Results and Discussion}
\vspace{-1ex}
The resulting Spearman correlation coefficients ($\rho$) and corresponding p-values are illustrated in Fig.~\ref{fig:correlation_results} to analyze age-wise, gender-wise, and location specific trends in the acoustic-spirometric relationship. The top row represents age-group results (20-30, 30-40, 40-50, and 50-60 years), while the bottom row shows correlations for all participants (n = 141) and separately for male and female participants. The x-axis represents the analyzed frequency bands (0-800, 100-200, 200-400, and 400-800 Hz), with bars corresponding to the posterior auscultation locations LL, LU, RU, and RL. The y-axis represents the Spearman correlation coefficient ($\rho$) between the subject-level median E/I ratio and FEV$_1$/FVC. A star ($\star$) denotes the auscultation location with the highest statistically significant correlation (p $<$ 0.05) within each frequency band.
\vspace{-1.5ex}
\subsubsection{Lower to Mid Frequency Bands (100-200 Hz and 200-400 Hz)}
\vspace{-1ex}
The lower-to-mid frequency bands (100-200 Hz and 200-400 Hz) consistently showed stronger and statistically significant correlations compared to the higher-frequency band (400-800 Hz) and the broadband 0-800 Hz range across age groups and genders. This observation aligns with the findings of Shimoda et al. and Malmberg et al., who reported that E/I values in the 100-400 Hz range are more strongly correlated with spirometry measures than other frequency components \cite{shimoda2017lung, shimoda2017lungf, malmberg1995significant}.

Shimoda et al. further noted that very low-frequency components (0-100 Hz) are more prone to physiological noise, such as cardiac and muscle sounds, and are less consistently correlated with airflow obstruction \cite{shimoda2017lungf}. Similarly, the 400-800 Hz band demonstrated weaker associations, likely due to the reduced dominance of airflow-related acoustic energy at higher frequencies. Consequently, the broadband 0-800 Hz range, which includes both low-frequency noise (0-100 Hz) and higher-frequency components (400-800 Hz), may show reduced airflow-specific spectral characteristics, leading to weaker associations with spirometric parameters. Together, these findings suggest that the 100-400 Hz region captures airflow-dependent acoustic changes more reliably and is particularly sensitive to expiratory airflow limitation in asthma.
\vspace{-1.5ex}
\subsubsection{All Age Groups and Genders}
\vspace{-1ex}
When analyzed across all participants, the Left Lower (LL) site showed stronger correlations with FEV$_1$/FVC compared to other posterior locations in the 100-200 Hz and 0-800 Hz bands. In the 200-400 Hz band, correlations at the Left Lower and Left Upper (LU) sites were comparable, whereas in the higher-frequency band (400-800 Hz), the Left Upper site demonstrated stronger correlations. These findings are consistent with Shimoda et al. \cite{shimoda2017lung}, who reported that the Left Lower posterior region often shows stronger associations with FEV$_1$/FVC compared to other posterior auscultation sites, particularly within the 100-400 Hz range.
\vspace{-1.5ex}
\subsubsection{Age-wise Analysis}
\vspace{-1ex}
In the 20-30 age group, the lower posterior sites, particularly the Left Lower region, showed the strongest and statistically significant correlations across most frequency bands, except in the 100-200 Hz band where the Right Lower site showed the highest correlation. In the 30--40 age group, the Left Lower site showed the strongest and statistically significant correlation in the 100--200 Hz band. For the remaining frequency bands, the Right Upper site showed the highest correlation in the 0--800 Hz broadband analysis, the Right Lower site showed the highest correlation in the 200--400 Hz band, and the Left Upper site showed the highest correlation in the 400--800 Hz band, although these correlations were not statistically significant. In the 40--50 age group, the Left Upper site exhibited the strongest correlations in the 200--400 Hz and 400--800 Hz bands, while the Left Lower site showed the strongest correlations in the 100--200 Hz and 0--800 Hz bands; all these correlations were statistically significant. Within the 50--60 age group, the Left Upper site exhibited the highest correlations across all frequency bands and showed statistically significant correlations in the 200--400 Hz and 400--800 Hz bands. These results are best interpreted as within-group trends, since no formal statistical comparisons were performed across age groups. Complete age-group-specific Spearman correlation coefficients, uncorrected p-values, and FDR-corrected p-values for all auscultation locations and frequency bands are provided in Appendix Table~\ref{tab:age_corr}.

The observed spatial patterns are consistent with known physiological and acoustic mechanisms associated with aging. Aging is associated with reduced lung elastic recoil and increased closing volume, leading to earlier small airway closure during expiration \cite{babb2000mechanism, turner1968elasticity}. As elastic recoil decreases, smaller airways in the lower lung regions become more prone to premature collapse, resulting in air trapping and reduced effective expiratory flow from the lung bases. Because FEV$_1$/FVC reflects dynamic airflow during forced expiration, early lower airway closure may reduce their contribution to the spirometric signal. Acoustic studies also show that regional lung sound distribution reflects ventilation and airflow patterns; for example, Lev et al. reported that increased ventilation (induced by higher positive end-expiratory pressure (PEEP)) redistributes lung sound energy toward the lower lung regions \cite{lev2009changes}. Therefore, if airflow from the lower regions declines with age, the E/I ratios measured at these sites may also decrease, causing upper posterior regions to contribute proportionally more to expiratory airflow and exhibit stronger correlations with FEV$_1$/FVC. The stronger correlations observed at upper posterior locations in older age groups are physiologically plausible and consistent with age-related increases in closing volume and loss of elastic recoil \cite{babb2000mechanism, turner1968elasticity}.
\vspace{-1.5ex}
\subsubsection{Gender-wise Analysis}
\vspace{-1ex}
The gender-wise analysis revealed distinct patterns. In males, the Left Lower site exhibited stronger and statistically significant correlations with FEV$_1$/FVC across all frequency bands compared to other auscultation locations. In contrast, in females, the Left Upper site showed stronger and statistically significant correlations across all frequency bands.

Although regional airflow redistribution with respect to gender has not been directly mapped in prior studies, physiological evidence indicates differences in airway diameter, lung size, and lung volume between males and females \cite{bellemare2003sex, dominelli2018sex, kim2011gender}. Males generally possess larger airways and greater lung volumes, which may influence airflow resistance and regional ventilation distribution, thereby plausibly contributing to the observed gender dependent variations in acoustic-spirometric relationships.

Together, these findings suggest that the anatomical regions most representative of dynamic expiratory airflow vary with both age and gender.

 Complete gender-specific Spearman correlation coefficients, uncorrected p-values, and FDR-corrected p-values for all auscultation locations and frequency bands are provided in Appendix Table~\ref{tab:gender_corr}.

 A supplementary two-way ANOVA performed on Fisher $z$-transformed correlation coefficients revealed a significant effect of gender and a significant age-by-gender interaction, suggesting that demographic factors may jointly influence the observed acoustic-spirometric relationships (Appendix Table~\ref{tab:anova_corr}).
 
\vspace{-1.5ex}
\section{Conclusion}
\vspace{-1ex}
In this work, we investigated the relationship between the expiratory-to-inspiratory (E/I) spectral power ratio and spirometric airflow limitation (FEV$_1$/FVC) across posterior auscultation locations, frequency bands, age groups, and genders in asthma participants. Correlation analysis showed stronger associations at lower posterior sites overall. The 100--200 Hz and 200--400 Hz bands consistently demonstrated significant correlations, indicating sensitivity to expiratory airflow limitation. Within individual age and gender groups, distinct spatial patterns were observed across auscultation locations. These
findings suggest that different auscultation locations may show
varying associations with airflow limitation within individual
demographic subgroups. Given that the observed correlations are moderate ($r = 0.2$--$0.4$), the findings should be viewed as exploratory evidence toward non-invasive respiratory assessment rather than a direct replacement for spirometry. Study limitations and considerations for interpreting the results are discussed in Appendix ~\ref{sec:limitations}. As part of future work, we plan to develop age-, gender-, and location-specific asthma classification models using machine learning to improve automated assessment of airflow obstruction.

\section{Acknowledgement}
The authors thank everyone involved in participant recruitment, data collection, annotation and discussion related to this work. Authors also thank the Department of Science and Technology (DST), Govt. of India for their support in this work.

⁠\section{Generative AI Use Disclosure} 
All authors are responsible and accountable for the work and content of this paper and consent to its submission. No generative AI tool is listed as an author. We used generative AI tools only for language editing and polishing (e.g., improving clarity and grammar) and not for producing any significant part of the manuscript. The study design, experiments, analysis, and conclusions are entirely the authors’ work, and all content was reviewed by the authors.

\bibliographystyle{IEEEtran}
\bibliography{mybib}

\clearpage
\appendix
\onecolumn
\section{Appendix}

\subsection{Welch's t-test Analysis of FEV$_1$/FVC Across Age Groups}

To assess whether differences in lung function severity could potentially influence the age-stratified correlation analysis, we performed pairwise Welch's t-tests on FEV$_1$/FVC values across the four age groups. Welch's t-test was selected because it does not assume equal variances and is appropriate for groups with unequal sample sizes.

The age-group statistics reported in Table 1 (here Table ~\ref{tab:age_fev1fvc}) of the main manuscript are reproduced below:

\begin{center}
\captionof{table}{Summary of FEV$_1$/FVC values across age groups.}
\label{tab:age_fev1fvc}

\begin{tabular}{lcc}
\toprule
Age Group & Participants (M + F) & FEV$_1$/FVC (\%) \\
\midrule
20--30 & 38 (20 + 18) & 83.30 $\pm$ 8.67 \\
30--40 & 42 (18 + 24) & 83.57 $\pm$ 9.60 \\
40--50 & 43 (21 + 22) & 79.37 $\pm$ 12.13 \\
50--60 & 18 (7 + 11) & 86.33 $\pm$ 7.40 \\
\bottomrule
\end{tabular}
\end{center}

Pairwise Welch's t-test results are shown in Table~\ref{tab:welch}.

\begin{center}
\captionof{table}{Pairwise Welch's t-test comparisons of FEV$_1$/FVC across age groups.}
\label{tab:welch}

\begin{tabular}{lc}
\toprule
Comparison & p-value \\
\midrule
20--30 vs 30--40 & 0.8934 \\
20--30 vs 40--50 & 0.0921 \\
20--30 vs 50--60 & 0.1793 \\
30--40 vs 40--50 & 0.0803 \\
30--40 vs 50--60 & 0.2343 \\
40--50 vs 50--60 & 0.0085 \\
\bottomrule
\end{tabular}
\end{center}

The results indicate that most age groups do not differ significantly in their FEV$_1$/FVC distributions ($p > 0.05$). The only statistically significant difference was observed between the 40--50 and 50--60 age groups ($p = 0.0085$), where the 50--60 group exhibited a higher mean FEV$_1$/FVC value (86.33\%) than the 40--50 group (79.37\%). Thus, pairwise comparisons show no significant differences between most age groups, except between the 40--50 and 50--60 groups. Therefore, we interpret the observed correlations as within-group associations rather than purely age-driven effects.

\subsection{Two-Way ANOVA Analysis of Age and Gender Effects}

To assess whether the observed age- and gender-specific correlation patterns may be influenced by interactions between demographic factors, a two-way ANOVA was performed using Fisher $z$-transformed correlation coefficients, with age group and gender as fixed factors.

\begin{center}
\captionof{table}{Two-way ANOVA results for Fisher $z$-transformed correlation coefficients.}
\label{tab:anova_corr}

\begin{tabular}{lccc}
\toprule
Source & F & p-value & Partial $\eta^2$ \\
\midrule
Gender & 8.53 & 0.0075 & 0.262 \\
Age Group & 1.83 & 0.1687 & 0.186 \\
Gender $\times$ Age Group & 4.57 & 0.0115 & 0.363 \\
\bottomrule
\end{tabular}

\vspace{0.5ex}

\footnotesize
\textit{Note:} $F$ denotes the ANOVA test statistic, and partial $\eta^2$ represents the corresponding effect size.
\end{center}

The analysis revealed a significant main effect of gender ($p = 0.0075$), indicating differences in the relationship between E/I ratio and FEV$_1$/FVC between male and female participants. No significant main effect of age group was observed ($p = 0.1687$). However, a significant age-by-gender interaction was detected ($p = 0.0115$), suggesting that the association between acoustic features and airflow limitation varies across demographic subgroups and cannot be explained by age or gender alone.

These findings support a cautious interpretation of the observed spatial patterns and indicate that age- and gender-related effects may interact in influencing the relationship between respiratory acoustics and spirometric airflow limitation.

\subsection{FDR-Corrected Statistical Significance Analysis}

To account for multiple comparisons, the Benjamini--Hochberg False Discovery Rate (FDR) correction was applied across all statistical tests reported in the study. Figure ~\ref{fig:without_fdr} presents the original significance map based on uncorrected $p$-values, while Figure ~\ref{fig:with_fdr} presents the corresponding results after FDR correction.

The significance markings were updated accordingly. A star ($\star$) indicates statistically significant correlations ($p < 0.05$ for the original analysis and FDR-corrected $p < 0.05$ for the corrected analysis), while a dot ($\cdot$) indicates the highest correlation among the four auscultation locations within a given frequency band. Unlike the original figure, where only the highest significant correlation within a frequency band was highlighted, the revised figures mark all statistically significant correlations, allowing clearer distinction between significant and non-significant results.

Although the number of statistically significant comparisons decreased following FDR correction, the principal spatial patterns remained largely unchanged. In particular, the dominant location-specific trends observed within individual age and gender groups were preserved after correction.

\begin{center}
\includegraphics[width=\linewidth]{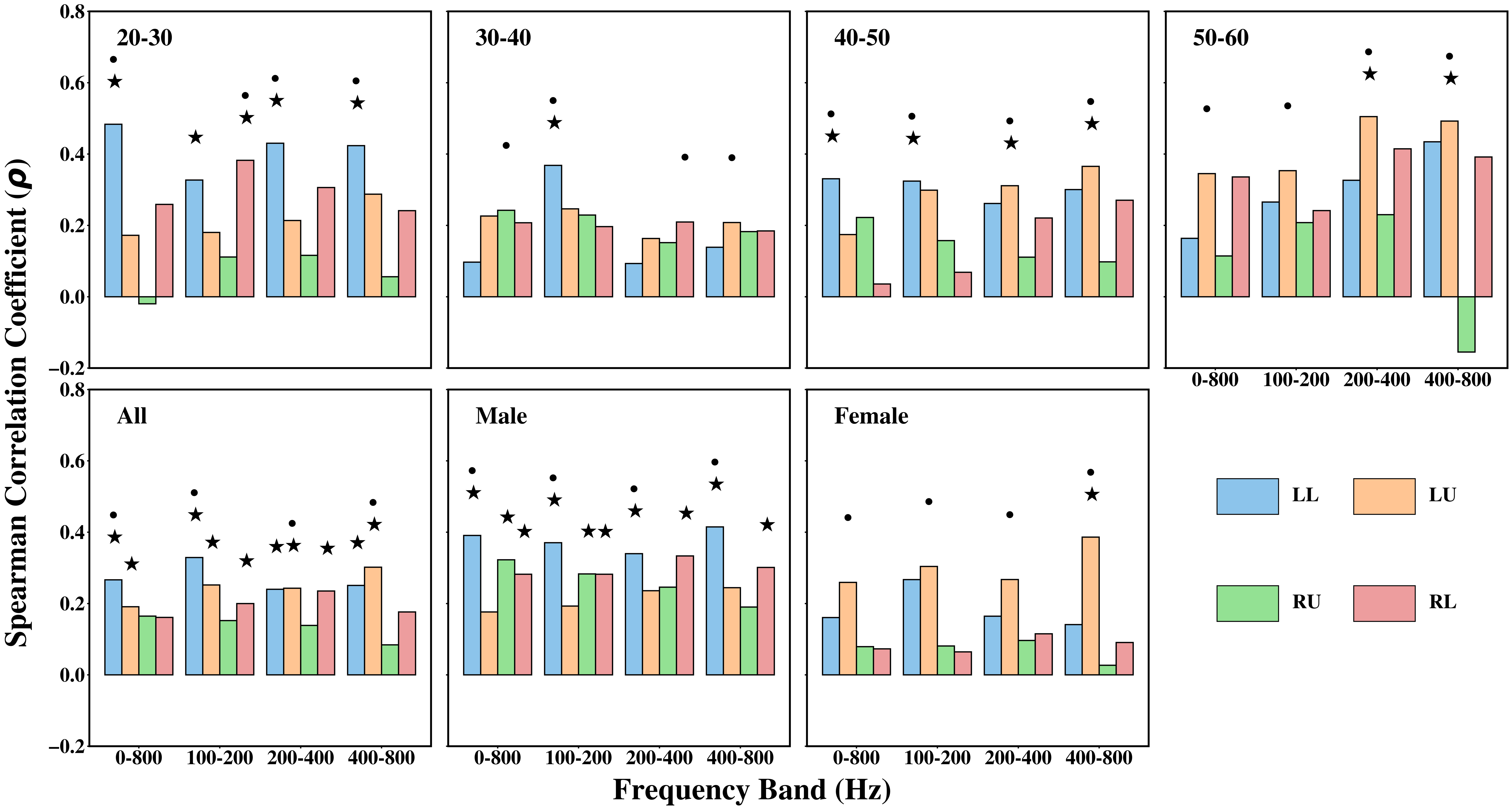}

\captionof{figure}{Original significance map based on uncorrected $p$-values. A star ($\star$) indicates statistical significance ($p<0.05$), while a dot ($\cdot$) indicates the highest correlation among locations within a frequency band.}
\label{fig:without_fdr}
\end{center}

\vspace{0.5cm}

\begin{center}
\includegraphics[width=\linewidth]{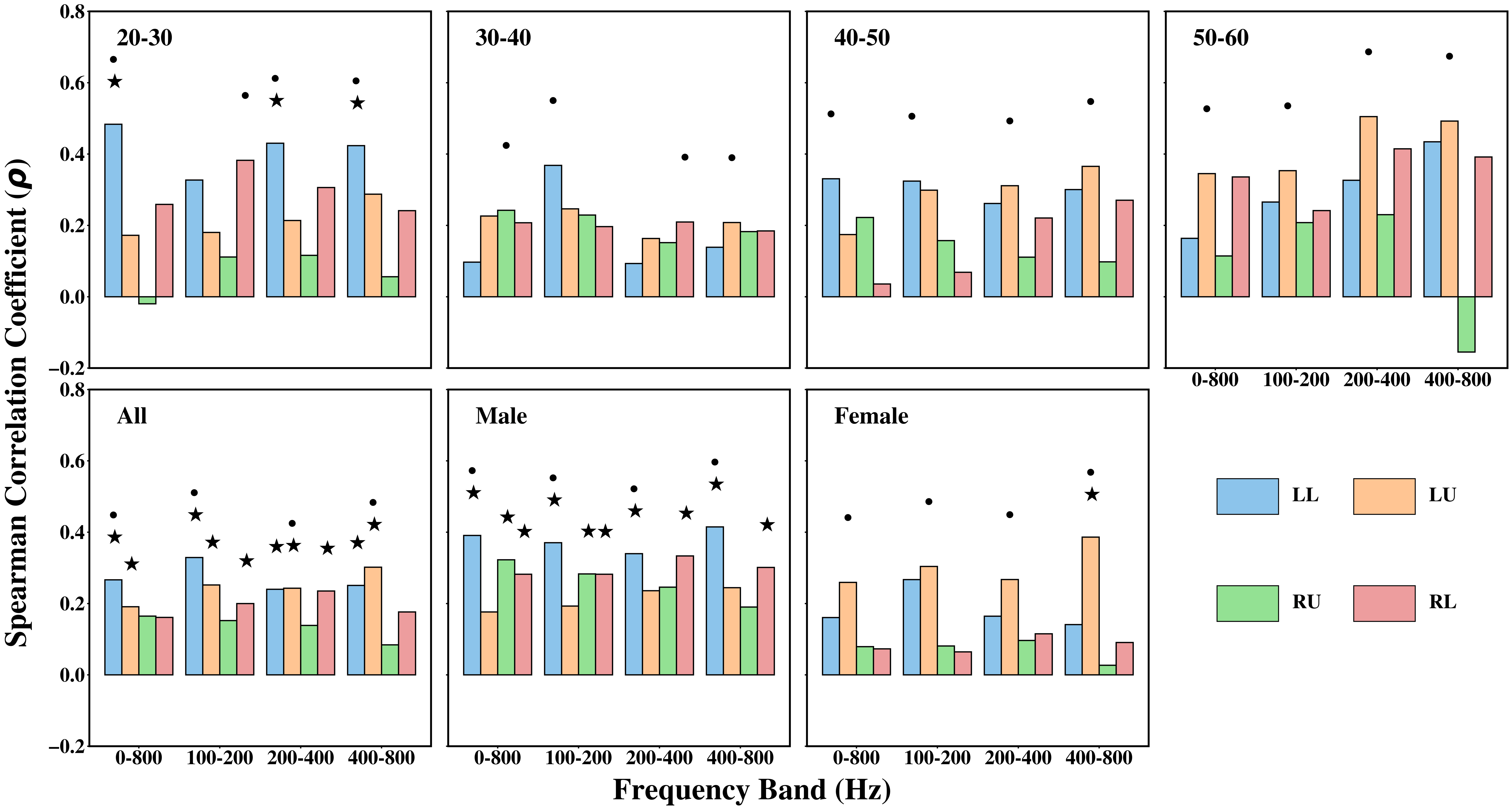}

\captionof{figure}{Significance map after Benjamini--Hochberg FDR correction. A star ($\star$) indicates statistical significance based on FDR-corrected $p$-values ($q<0.05$), while a dot ($\cdot$) indicates the highest correlation among locations within a frequency band.}
\label{fig:with_fdr}
\end{center}

\subsection{Complete Correlation Results}
Table \ref{tab:age_corr} and Table \ref{tab:gender_corr} provides the complete set of Spearman correlation coefficients, uncorrected p-values, and Benjamini--Hochberg FDR corrected p-values for all age-group and gender-group analyses.

\small

\begin{longtable}{lllrrr}
\caption{Age-wise Spearman correlation analysis.}
\label{tab:age_corr}\\

\toprule
Age Group & Location & Band & Spearman $r$ & $p$-value & FDR-corrected $p$ \\
\midrule
\endfirsthead

\multicolumn{6}{c}%
{{\tablename\ \thetable{} -- continued from previous page}}\\
\toprule
Age Group & Location & Band & Spearman $r$ & $p$-value & FDR-corrected $p$ \\
\midrule
\endhead

\midrule
\multicolumn{6}{r}{Continued on next page}\\
\endfoot

\bottomrule
\endlastfoot

20--30 & LL & 0--800 & 0.4836 & 0.0021 & 0.0302 \\
20--30 & LL & 100--200 & 0.3272 & 0.0449 & 0.1324 \\
20--30 & LL & 200--400 & 0.4304 & 0.0070 & 0.0489 \\
20--30 & LL & 400--800 & 0.4235 & 0.0081 & 0.0501 \\
20--30 & LU & 0--800 & 0.1722 & 0.3013 & 0.4022 \\
20--30 & LU & 100--200 & 0.1802 & 0.2789 & 0.3857 \\
20--30 & LU & 200--400 & 0.2137 & 0.1976 & 0.2950 \\
20--30 & LU & 400--800 & 0.2876 & 0.0799 &	0.1792 \\
20--30 & RU	& 0--800 & -0.0197 & 0.9065 & 0.9065 \\
20--30 & RU & 100--200 & 0.1115 & 0.5050 & 0.5713 \\
20--30 & RU & 200--400 & 0.1160 & 0.4879 & 0.5648 \\
20--30 & RU & 400--800 & 0.0562 & 0.7375 & 0.7578 \\
20--30 & RL & 0--800 & 0.2589 & 0.1166 & 0.2331 \\
20--30 & RL & 100--200 & 0.3826 & 0.0178 & 0.0829 \\
20--30 & RL & 200--400 & 0.3062 & 0.0615 & 0.1498 \\
20--30 & RL & 400--800 & 0.2412 & 0.1446 & 0.2644 \\

30--40 & LL & 0--800 & 0.0970 & 0.5410 & 0.5882 \\
30--40 & LL & 100--200 & 0.3684 & 0.0164 & 0.0829 \\
30--40 & LL & 200--400 & 0.0932 & 0.5571 & 0.5942 \\
30--40 & LL & 400--800 & 0.1387 & 0.3809 & 0.4688 \\
30--40 & LU & 0--800 & 0.2264 & 0.1494 & 0.2644 \\
30--40 & LU & 100--200 & 0.2464 & 0.1157 & 0.2331 \\
30--40 & LU & 200--400 & 0.1632 & 0.3016 & 0.4022 \\
30--40 & LU & 400--800 & 0.2081 & 0.1860 & 0.2836 \\
30--40 & RU & 0--800 & 0.2424 & 0.1219 & 0.2354 \\
30--40 & RU & 100--200 & 0.2290 & 0.1445 & 0.2644 \\
30--40 & RU & 200--400 & 0.1516 & 0.3377 & 0.4250 \\
30--40 & RU & 400--800 & 0.1826 & 0.2472 & 0.3505 \\
30--40 & RL & 0--800 & 0.2075 & 0.1874 & 0.2836 \\
30--40 & RL & 100--200 & 0.1966 & 0.2121 & 0.3126 \\
30--40 & RL & 200--400 & 0.2096 & 0.1828 & 0.2836 \\
30--40 & RL & 400--800 & 0.1846 & 0.2419 & 0.3474 \\

40--50 & LL & 0--800 & 0.3308 & 0.0303 & 0.1059 \\
40--50 & LL & 100--200 & 0.3242 & 0.0339 & 0.1117 \\
40--50 & LL & 200--400 & 0.2615 & 0.0903 & 0.1945 \\
40--50 & LL & 400--800 & 0.3004 & 0.0503 & 0.1350 \\
40--50 & LU & 0--800 & 0.1745 & 0.2631 & 0.3684 \\
40--50 & LU & 100--200 & 0.2987 & 0.0517 & 0.1350 \\
40--50 & LU & 200--400 & 0.3112 & 0.0422 & 0.1278 \\
40--50 & LU & 400--800 & 0.3655 & 0.0159 & 0.0829 \\
40--50 & RU & 0--800 & 0.2222 & 0.1520 & 0.2644 \\
40--50 & RU & 100--200 & 0.1573 & 0.3137 & 0.4134 \\
40--50 & RU & 200--400 & 0.1110 & 0.4787 & 0.5643 \\
40--50 & RU & 400--800 & 0.0978 & 0.5326 & 0.5867 \\
40--50 & RL & 0--800 & 0.0357 & 0.8203 & 0.8277 \\
40--50 & RL & 100--200 & 0.0686 & 0.6622 & 0.6867 \\
40--50 & RL & 200--400 & 0.2208 & 0.1548 & 0.2644 \\
40--50 & RL & 400--800 & 0.2706 & 0.0792 & 0.1792 \\

50--60 & LL & 0--800 & 0.1638 & 0.5160 & 0.5779 \\
50--60 & LL & 100--200 & 0.2654 & 0.2871 & 0.3921 \\
50--60 & LL & 200--400 & 0.3266 & 0.1859 & 0.2836 \\
50--60 & LL & 400--800 & 0.4344 & 0.0716 & 0.1692 \\
50--60 & LU & 0--800 & 0.3453 & 0.1605 & 0.2644 \\
50--60 & LU & 100--200 & 0.3536 & 0.1501 & 0.2644 \\
50--60 & LU & 200--400 & 0.5049 & 0.0326 & 0.1106 \\
50--60 & LU & 400--800 & 0.4925 & 0.0379 & 0.1178 \\
50--60 & RU & 0--800 & 0.1144 & 0.6620 & 0.6867 \\
50--60 & RU & 100--200 & 0.2079 & 0.4234 & 0.5099 \\
50--60 & RU & 200--400 & 0.2300 & 0.3745 & 0.4660 \\
50--60 & RU & 400--800 & -0.1550 & 0.5525 & 0.5942 \\
50--60 & RL & 0--800 & 0.3359 & 0.1729 & 0.2767 \\
50--60 & RL & 100--200 & 0.2416 & 0.3342 & 0.4250 \\
50--60 & RL & 200--400 & 0.4147 & 0.0870 & 0.1911 \\
50--60 & RL & 400--800 & 0.3919 & 0.1077 & 0.2234 \\
\end{longtable}

\scriptsize
\small
\begin{longtable}{lllrrr}
\caption{Gender-wise Spearman correlation analysis between E/I ratio and FEV$_1$/FVC.}
\label{tab:gender_corr}\\

\toprule
Group & Location & Band & Spearman $r$ & $p$-value & FDR-corrected $p$ \\
\midrule
\endfirsthead

\multicolumn{6}{c}%
{{\tablename\ \thetable{} -- continued from previous page}}\\
\toprule
Group & Location & Band & Spearman $r$ & $p$-value & FDR-corrected $p$ \\
\midrule
\endhead

\midrule
\multicolumn{6}{r}{Continued on next page}\\
\endfoot

\bottomrule
\endlastfoot

All & LL & 0--800 & 0.2664 & 0.0014 & 0.0262 \\
All & LL & 100--200 & 0.3291 & 0.0001 & 0.0076 \\
All & LL & 200--400 & 0.2400 & 0.0042 & 0.0388 \\
All & LL & 400--800 & 0.2509 & 0.0027 & 0.0302 \\
All & LU & 0--800 & 0.1910 & 0.0233 & 0.0869 \\
All & LU & 100--200 & 0.2521 & 0.0026 & 0.0302 \\
All & LU & 200--400 & 0.2430 & 0.0037 & 0.0375 \\
All & LU & 400--800 & 0.3018 & 0.0003 & 0.0155 \\
All & RU & 0--800 & 0.1647 & 0.0518 & 0.1350 \\
All & RU & 100--200 & 0.1523 & 0.0725 & 0.1692 \\
All & RU & 200--400 & 0.1387 & 0.1021 & 0.2157 \\
All & RU & 400--800 & 0.0843 & 0.3221 & 0.4189 \\
All & RL & 0--800 & 0.1610 & 0.0564 & 0.1404 \\
All & RL & 100--200 & 0.1998 & 0.0175 & 0.0829 \\
All & RL & 200--400 & 0.2351 & 0.0050 & 0.0420 \\
All & RL & 400--800 & 0.1763 & 0.0365 & 0.1167 \\

Male & LL & 0--800 & 0.3910 & 0.0012 & 0.0262 \\
Male & LL & 100--200 & 0.3705 & 0.0022 & 0.0302 \\
Male & LL & 200--400 & 0.3398 & 0.0053 & 0.0420 \\
Male & LL & 400--800 & 0.4149 & 0.0005 & 0.0173 \\
Male & LU & 0--800 & 0.1764 & 0.1564 & 0.2644 \\
Male & LU & 100--200 & 0.1929 & 0.1207 & 0.2354 \\
Male & LU & 200--400 & 0.2360 & 0.0564 & 0.1404 \\
Male & LU & 400--800 & 0.2444 & 0.0480 & 0.1350 \\
Male & RU & 0--800 & 0.3226 & 0.0088 & 0.0517 \\
Male & RU & 100--200 & 0.2830 & 0.0223 & 0.0863 \\
Male & RU & 200--400 & 0.2460 & 0.0482 & 0.1350 \\
Male & RU & 400--800 & 0.1901 & 0.1293 & 0.2455 \\
Male & RL & 0--800 & 0.2823 & 0.0217 & 0.0863 \\
Male & RL & 100--200 & 0.2822 & 0.0217 & 0.0863 \\
Male & RL & 200--400 & 0.3334 & 0.0062 & 0.0465 \\
Male & RL & 400--800 & 0.3012 & 0.0140 & 0.0784 \\

Female & LL & 0--800 & 0.1606 & 0.1686 & 0.2737 \\
Female & LL & 100--200 & 0.2671 & 0.0205 & 0.0863 \\
Female & LL & 200--400 & 0.1646 & 0.1583 & 0.2644 \\
Female & LL & 400--800 & 0.1411 & 0.2274 & 0.3307 \\
Female & LU & 0--800 & 0.2593 & 0.0247 & 0.0892 \\
Female & LU & 100--200 & 0.3041 & 0.0080 & 0.0501 \\
Female & LU & 200--400 & 0.2674 & 0.0204 & 0.0863 \\
Female & LU & 400--800 & 0.3863 & 0.0006 & 0.0173 \\
Female & RU & 0--800 & 0.0791 & 0.4999 & 0.5713 \\
Female & RU & 100--200 & 0.0811 & 0.4891 & 0.5648 \\
Female & RU & 200--400 & 0.0965 & 0.4103 & 0.4995 \\
Female & RU & 400--800 & 0.0271 & 0.8178 & 0.8277 \\
Female & RL & 0--800 & 0.0729 & 0.5344 & 0.5867 \\
Female & RL & 100--200 & 0.0644 & 0.5828 & 0.6158 \\
Female & RL & 200--400 & 0.1151 & 0.3254 & 0.4189 \\
Female & RL & 400--800 & 0.0907 & 0.4389 & 0.5230 \\

\end{longtable}

\normalsize

\subsection{Study Limitations}
\label{sec:limitations}

Several limitations of the present study should be considered when interpreting the results. First, the analysis was conducted exclusively on asthma participants, and no healthy control group was included in the correlation analysis. Consequently, the observed relationships between E/I ratio and FEV$_1$/FVC cannot be directly generalized to healthy populations.

Second, recordings were obtained during a single session, and test--retest reliability was not evaluated. Future longitudinal studies are required to assess the stability and reproducibility of the proposed acoustic measures.

Third, some demographic subgroups contained relatively small sample sizes, particularly the 50--60 age group. As a result, subgroup-specific correlation patterns should be interpreted cautiously and viewed as exploratory observations.

Fourth, the physiological processes underlying passive tidal breathing and forced spirometric maneuvers are not identical. While significant associations were observed between E/I ratio and FEV$_1$/FVC, the acoustic measurements and spirometric indices reflect related but distinct aspects of respiratory function.

Finally, all analyses were performed using a single dataset collected under a common acquisition protocol. External validation on independent cohorts and recording environments will be necessary to establish the generalizability of the reported findings.

\end{document}